\documentclass[fleqn,usenatbib]{mnras}

\usepackage{newtxtext,newtxmath}
\usepackage[T1]{fontenc}

\DeclareRobustCommand{\VAN}[3]{#2}
\let\VANthebibliography\thebibliography
\def\thebibliography{\DeclareRobustCommand{\VAN}[3]{##3}\VANthebibliography}

\usepackage{graphicx}	% Including figure files
\usepackage{amsmath}	% Advanced maths commands
\usepackage{CJKutf8}    % Chinese characters
\usepackage{color,soul} % highlighting (temp)
\usepackage{verbatim}   % Comments (temp)
\usepackage{pdflscape}  % For landscape
\usepackage[shortlabels]{enumitem}   % For enumerate label
\PassOptionsToPackage{hyphens}{url}
\usepackage{tikz,xcolor,hyperref} % For ORCID
\graphicspath{ {fig/} }
\usepackage{booktabs} % For \toprule, \midrule and \bottomrule
\usepackage{siunitx} % Formats the units and values
\usepackage{pgfplotstable} % Generates table from .csv
\pgfplotsset{compat=1.18}
\newcommand{\mgii}{Mg\,\textsc{ii}\ }

\newcommand{\lya}{Ly\,\textsc{$\alpha$}}

\newcommand{\lrf}{$\lambda_{\rm rf}$}

\newcommand{\tbrk}{$t_{\rm brk}$}

\newcommand{\glin}{$\gamma_{\rm lin}$}
\newcommand{\gsh}{$\gamma_{\rm 1,pw}$}
\newcommand{\glo}{$\gamma_{\rm 2,pw}$}

\newcommand{\mbh}{$M_{\rm BH}$}
\newcommand{\lbol}{$L_{\rm bol}$}

\newcommand{\lwv}{$L_{3000}$}

\newcommand{\sigmed}{$\sigma_{\rm mag,med}$}
\newcommand{\sigsq}{$\langle \sigma^2 \rangle$}
\newcommand{\dm}{$\langle \Delta m \rangle$}
\newcommand{\dmsq}{$\langle \Delta m \rangle^2$}

\setlist[enumerate]{
  labelsep=8pt,
  labelindent=0\parindent,
  itemindent=0pt,
  leftmargin=*,
  before=\setlength{\listparindent}{-\leftmargin},
}

\definecolor{lime}{HTML}{A6CE39}
\DeclareRobustCommand{\orcidicon}{%
    \begin{tikzpicture}
    \draw[lime, fill=lime] (0,0) 
    circle [radius=0.16] 
    node[white] {{\fontfamily{qag}\selectfont \tiny ID}};
    \draw[white, fill=white] (-0.0625,0.095) 
    circle [radius=0.007];
    \end{tikzpicture}
    \hspace{-2mm}
}

\newcommand{\orcidJJT}{\href{https://orcid.org/0000-0002-1860-0886}{\orcidicon}}
\newcommand{\orcidChrisW}{\href{https://orcid.org/0000-0002-4569-016X}{\orcidicon}}
\newcommand{\orcidJT}{\href{https://orcid.org/0000-0003-2858-9657}{\orcidicon}}

\title[Quasar VSF at Short Timescales]{The Variability Structure Function of the Highest-Luminosity Quasars on Short Timescales}

\author[J.-J. Tang et al.]
{Ji-Jia Tang$^{1,2}$\orcidJJT \thanks{E-mail: ji-jia.tang@anu.edu.au, ji-jia.tang@phys.ntu.edu.tw},
Christian Wolf$^{1,3}$\orcidChrisW \thanks{E-mail: christian.wolf@anu.edu.au},
John Tonry$^4$\orcidJT \\
$^1$Research School of Astronomy and Astrophysics, Australian National University, Cotter Road, Weston Creek ACT 2611, Australia \\
$^2$Graduate Institute of Astrophysics and Department of Physics, National Taiwan University, No. 1, Sec. 4 Roosevelt Road, Taipei 10617, Taiwan \\
$^3$Centre for Gravitational Astrophysics (CGA), Australian National University, Building 38 Science Road, Acton ACT 2601, Australia \\
$^4$Institute for Astronomy, University of Hawaii, 2680 Woodlawn Drive, Honolulu, HI 96822-1897, U.S.A. \\
}

\date{Accepted XXX. Received YYY; in original form ZZZ}

\pubyear{2024}

\begin{document}
\label{firstpage}
\pagerange{\pageref{firstpage}--\pageref{lastpage}}
\maketitle

\begin{abstract}
The stochastic photometric variability of quasars is known to follow a random-walk phenomenology on emission timescales of months to years. Some high-cadence restframe optical monitoring in the past has hinted at a suppression of variability amplitudes on shorter timescales of a few days or weeks, opening the question of what drives the suppression and how it might scale with quasar properties. Here, we study a few thousand of the highest-luminosity quasars in the sky, mostly in the luminosity range of \lbol$=[46.4, 47.3]$ and redshift range of $z=[0.7, 2.4]$. We use a dataset from the NASA/ATLAS facility with nightly cadence, weather permitting, which has been used before to quantify strong regularity in longer-term restframe-UV variability. As we focus on a careful treatment of short timescales across the sample, we find that a linear function is sufficient to describe the UV variability structure function. Although the result can not rule out the existence of breaks in some groups completely, a simpler model is usually favoured under this circumstance. In conclusion, the data is consistent with a single-slope random walk across restframe timescales of $\Delta t=[10, 250]$~days. 
\end{abstract}


%%%%%%%%%%%%%%%%%%%%%%%%%%%%%%%%%%%%%%%%%%%%%%%%%%

%%%%%%%%%%%%%%%%% BODY OF PAPER %%%%%%%%%%%%%%%%%%

\defcitealias{TWT}{Paper I}
\defcitealias{incl}{Paper II}

\section{Introduction}
\label{intro}

In the last two decades, the field of quasar variability has made tremendous progress due to the availability of large time domain surveys \citep[e.g.][]{Mo14, Ca17, Li18, Su21}. These data sets are hoped to reveal correlations between variability behaviour and physical quasar properties to reveal the underlying physics behind the variability of quasar accretion discs \citep[e.g.][]{Ulr97, Pe01, Pad17, Cack21}. The variability could be caused by the turbulence that is driven by magneto rotational instability \citep[MRI;][]{BH91}, and in addition the disc will respond to variable heating from the X-ray corona near the inner disc, an effect known as reprocessing or lamppost effect \citep[e.g.][]{Cl92, Shap14, HS20, Cack21}.

A common technique to study quasar variability in the UV/optical domain is structure function \citep[SF;][]{Hugh92} analysis: this is suitable for analysing observed light curves (LC) with non-uniform sampling, which is often a natural consequence of weather and other constraints on observing. The variability SF of an LC shows how observed brightness changes on average as a function of time interval. The detailed mathematical background of SF analysis is described very well by \citet{Koz16b}.

If the brightness change of the quasar is truly stochastic, i.e., there is no correlation of a present change with any of the past changes, the behaviour can be described as the random walk (RW). According to the random walk model, the power-law slope of the SF is expected to be $0.5$ \citep{Ke09}, indicating the source of variability is caused by disc instability \citep{Take95, Kw98}. This $0.5$ slope is confirmed by observations \citep[e.g.][]{Ren06, Ca17, Li18, TWT} on timescales $\gtrapprox$100 days. Likewise, studies that expressed the variability as a power spectral densities (PSD) show a slope of $-2$ in the PSD, which is mathematically equivalent to a slope of $0.5$ in the SF \citep[e.g.][]{Giv99, Ke09}.

However, \citet{Mu11} found the slope on timescales of a few days seems to be significantly steeper than 0.5, indicating the existence of a break. Several studies that cover a broad range of timescales have found breaks on short timescales \citep[e.g.][]{Edel14, Ke14, Simm16, Sm18, St22}, motivating attempts to explain this feature in the structure function \citep[e.g.][]{Sun20a, Ta20} or to model quasar LCs including this effect \citep[e.g.][]{Ke14, Ka17, More19, Yu22}. However, other studies that also probe short timescales find no evidence of breaks \citep[e.g.][]{Mo14, Ca17}, casting doubt on whether there is any slope change or not.

The damped random walk \citep[DRW;][]{Ke09} model predicts that the variability at the timescales longer than the random walk regime becomes uncorrelated with the timescale, producing a flatten slope $0$ in SF. The breaking timescale indicating this turnover is often referred to as characteristic timescale or de-correlation timescale. To clarify, we focus on the break shorter than this break in this work.

Previously, we have used the NASA/Asteroid Terrestrial-impact Last Alert System \citep[ATLAS;][]{To18a} survey to study the quasar variability, as it has proven to be one of few very rich data sets. \citet[][hereafter \citetalias{TWT}]{TWT} performed the SF analysis on optical ATLAS LCs of $\sim 5\,000$ brightest quasars and showed that the random-walk relation is universal in variability amplitude when time is expressed in units of orbital or thermal timescale of the emitting portion of the accretion disc, seemingly independent of physical parameters that are not encapsulated in the disc time scales. This has emphasized the similarity between quasar accretion discs and pointed to evidence that MRI may be related to the root cause of the variability. While that work has shown breaks on short timescales, its analysis was focused on the longer-term random walk. In this work, we aim to deepen our analysis of the short time scales of the same light curves from NASA/ATLAS. In Section~\ref{sec:data_samp}, we provide a description of the sample and data. We outline the method employed to calculate and fit the SF in Section~\ref{sec:meth} and show the result in Section~\ref{sec:resul}. Throughout our analysis, we adopt a flat $\Lambda$CDM cosmology with $\Omega_\Lambda = 0.7$ and $H_0 = 70$~km~s$^{-1}$~Mpc$^{-1}$. Additionally, we use Vega magnitudes for data from {\it Gaia} and AB magnitudes for data from NASA/ATLAS and for absolute magnitude estimates.

\section{Data and sample}
\label{sec:data_samp}

We use the light curves from the analysis of the structure function in \citetalias{TWT} but reduce the sample somewhat to objects with good-quality spectral fitting and black-holes mass estimates from \citet{Rak20}. \citetalias{TWT} had selected 6\,163 spectroscopically confirmed, bright ({\it Gaia} magnitude $G_{\rm RP}<17.5$), redshift $0.5<z<3.5$, non-lensed\footnote{\href{https://research.ast.cam.ac.uk/lensedquasars/}{https://research.ast.cam.ac.uk/lensedquasars/}}\citep{Le19, Le23}, and isolated quasars from the Million Quasars Catalogue \citep[MILLIQUAS v7.1 update;][]{Flesch15} matched with the {\it Gaia} eDR3 catalogue \citep{gaia21}. These quasars have declination $\delta>-45^\circ$ to match the sky coverage of NASA/ATLAS and Galactic foreground reddening \citep{SFD98} of $E(B-V)_{\rm SFD}<0.15$. To avoid photometric contamination from nearby sources, the {\it Gaia} BpRp Excess Factor was limited to 1.3. \citetalias{TWT} obtained the LCs from the NASA/ATLAS database between 2015 to 2021 in two passbands, orange (6785\AA) and cyan (5330\AA), whose magnitudes are denoted by $m_{\rm o}$ and $m_{\rm c}$, respectively. They rejected quasars with noisy LCs, i.e., where the 90-percentile flux errors in the LCs were above $\sigma_{f_\nu, {\rm o}} > 150\mu$Jy or $\sigma_{f_\nu, {\rm c}} > 85\mu$Jy. 

\citetalias{TWT} rejected radio-loud quasars as their optical variability might be contaminated by processes outside of the accretion disc. They cross-matched the sample with catalogues from the Faint Images of the Radio Sky at Twenty-cm \citep[FIRST;][]{Be95} Survey, the NRAO VLA Sky Survey \citep[NVSS;][]{Co98}, and the Sydney University Molonglo Sky Survey \citep[SUMSS;][]{Ma03}. To account for the different spatial resolution of the radio surveys, they matched the radii between the {\it Gaia} coordinates and the FIRST, NVSS, and SUMSS coordinates with 3", 12", and 11", respectively. When multiple radio sources in NVSS or SUMSS were matched to one quasar, the one with the closest separation was used. The radio loudness criterion for matches with NVSS and SUMSS magnitudes ($t_{\rm NVSS/SUMSS}$) was $0.4\times(m_{\rm o}-t_{\rm NVSS/SUMSS})=1.3$, below which radio-detected objects were labelled as radio-intermediate. All quasars with matches in FIRST but not in NVSS were radio-intermediate, too. After that, 5\,315 quasars were left over.

\citetalias{TWT} cleaned the NASA/ATLAS LCs by excluding all observations with large errors of $\log(\sigma_{f_\nu, {\rm c}})>-4.17-0.10\times(m_{\rm c}-16.5)$ and $\log(\sigma_{f_\nu, {\rm o}})>-3.94-0.12\times(m_{\rm o}-16.5)$; and further removed outliers by comparing each measurement with other observations within $\pm$7 days and then using a $2\sigma$-clipping technique to reject spurious outliers. Observations were retained if they are isolated within $\pm$7 days. Altogether, 11\% of the data points are removed by the cleaning. They found that data at rest-frame wavelengths \lrf{} $\gtrapprox3000$\AA~do not follow the simple mean relation, and they also avoided wavelengths that were short enough to be contaminated by Ly$\alpha$ emission. After discarding LCs from affected passbands, they were left with 4\,724 quasars.

Here, we match the 4\,724 quasars with the \citet{Rak20} catalogue to obtain quasar properties derived from the SDSS DR14 spectral fitting. Using a separation radius within 1 arcsec, we find 3\,253 matches. We adopt the black hole mass, \mbh{}, measured from the \mgii line and calibrated with the \citet{VO09} relation in the \citet{Rak20} catalogue. There are 2\,750 quasars with good \mbh{} estimates, i.e. LOG\_MBH\_MGII\_VO09 > 0 and QUALITY\_MBH = 0. Among this sample, 13 of them have their cyan data contaminated by the strong  \lya{} line. After excluding them, we keep 2\,737 quasars for further analysis.

We adopt luminosities at rest-frame 3000\AA \ (\lwv) calculated in \citetalias{TWT}. They used {\it Gaia} $G_{\rm BP}$ and $G_{\rm RP}$ magnitudes to interpolate \lwv{} although at redshift $z>2$ this becomes a mild extrapolation beyond the coverage of the $G_{\rm RP}$ bandpass. They also correct for Milky Way \citep{SFD98} foreground dust extinction with bandpass coefficients $R_{\rm G_{BP}}=3.378$ and $R_{\rm G_{RP}}=2.035$ \citep{CV18}. They finally apply an offset of -0.07~dex after comparing their \lwv{} with the values in \citet{Rak20} that are derived from spectral decomposition into continuum and emission-line contributions. Bolometric luminosities, \lbol, were derived using $\log{(\text{\lbol}/\text{\lwv})}=\log{(f_{\rm BC} \times \lambda)}=\log{(5.15\times3000{\text \AA})}=4.189$, where $f_{\rm BC}$ is the bolometric correction factor \citep{Rich06b}.

\section{Methods}
\label{sec:meth}

\subsection{Variability structure functions}
\label{sec:VSF}

The SF analysis has been developed to quantify the observed variability in an LC, especially for the common cases of uneven sampling \citep{Hugh92}. In this work, we adopt the noise-corrected definition of variability amplitude from \citet{Cl96}:
\begin{equation}
	A=\sqrt{\frac{\pi}{2} \text{\dmsq} - \text{\sigsq}},   ~ ~ {\rm with} ~ ~
	\Delta m=|m_{\rm i}-m_{\rm j}|,
	\label{eq:sf}
\end{equation}
where $m_{\rm i}$ and $m_{\rm j}$ are any two observed apparent magnitudes and $\sigma$ is the noise due to magnitude error.

\begin{figure}
\begin{center}
\includegraphics[width=\columnwidth]{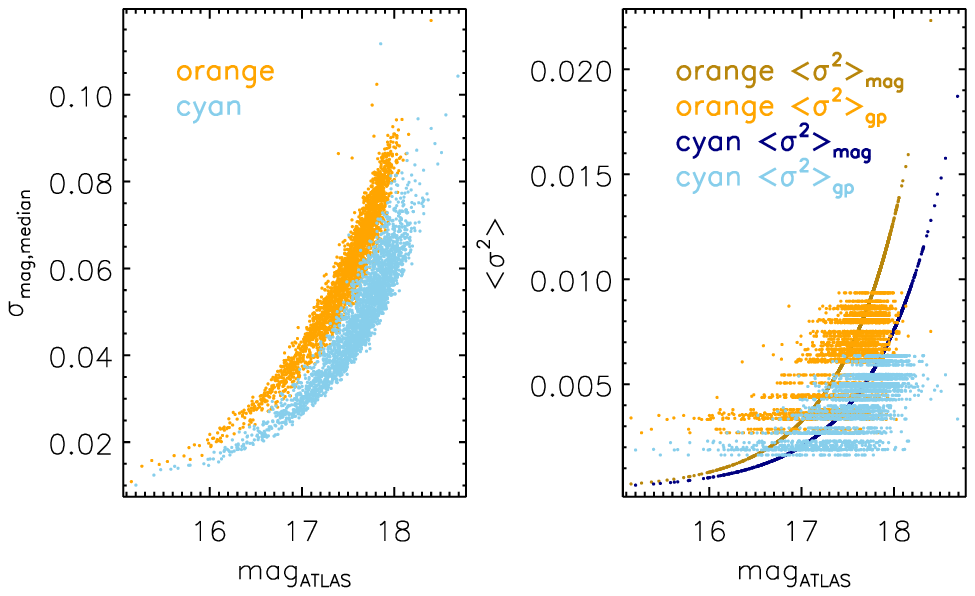}
\caption[The magnitude error and the comparison between \sigsq{} defined in two methods.]{Left: The median magnitude error (\sigmed) of the sample. Right: The comparison between \sigsq{} using the first definition (\sigsq$_{\rm mag}$) and the second definition (\sigsq$_{\rm gp}$).}
\label{fig:nlvsq}
\end{center}
\end{figure}

This work focuses on short-term variability, where an accurate handling of the noise is critical. Thus, we will be testing two different methods to estimate the mean error \sigsq{}. The first method follows the procedure in \citetalias{TWT}. Their \sigsq{} was determined as a function of observed magnitude, $m_{\rm obs}$, such that the SF for $\Delta t < 1$ day becomes $A\approx 0$ \citep{Koz16b}. This strategy was chosen because the typical amplitude of intraday variability in radio-quiet quasars is so small that it is within the uncertainty of the noise itself and will not noticeably affect the analysis of the random walk portion in the SF. They fitted a noise model to the ATLAS data using 
\begin{equation}
	\log(\text{\sigsq})\equiv\log{\left(\frac{\pi}{2}\text{\dmsq}\right)}=n_0+n_1 m_{\rm obs} ~,
	\label{eq:noise}
\end{equation}
for $\Delta t < 1$ day of their sample and found $(n_0, n_1)= (-12.411, 0.585)$ and $(-12.428, 0.573)$ for the orange and cyan passbands, respectively. Applying the same model for the 2\,750 quasars in this work, we find $(n_0, n_1)=(-12.683, 0.600)$ and $(-12.360, 0.569)$ for the orange and cyan passbands, respectively. For reference, the left panel of Fig.~\ref{fig:nlvsq} shows the median magnitude error, \sigmed, of those 2\,750 quasars. For $m_{\rm obs}=(16, 17.5)$, \sigmed$=(0.021, 0.059)$ for orange passband and $(0.016, 0.043)$ for cyan passband. They are comparable to the first definition of \sigsq{} after taking the squared value times $2$. After applying the noise subtraction with these parameters, we find that the term in the square-root of Eq.~\ref{eq:sf} is negative for about two thirds of the intraday magnitude pairs instead of the 50\% expected for pure noise without true variability. Thus, we choose a second definition of \sigsq{} following again the concept of setting the SF for $\Delta t < 1$ day to $A\approx 0$; however, this time, we do not try to fit a global magnitude-dependent function of the form above to the whole sample. Instead, we choose \sigsq{} to be the median value of the $(\pi/2)$\dmsq{} for $\Delta t < 1$ day in each sub-sample group defined in Sec.~\ref{sec:bin}, in order to force set the recovered variability amplitude to zero at short times. In the right panel of Fig.~\ref{fig:nlvsq}, we show the comparison between the \sigsq{} defined using these two methods. If this approach was inappropriate because of true intraday variability being present and inflating the signal, it would bias the inferred amplitude lower especially at lowest amplitudes, i.e., at shortest time scales; hence, such an effect would enhance an apparent suppression of short-term amplitude in the structure function, but it could not cause a bias towards reducing any suppression we find.

\begin{table}
  \begin{center}
    \caption{Boundaries of \lwv, \mbh, and number of quasars in each group.
    }
    \label{tab:bin}
    \begin{filecontents*}{Tab1.csv}
(0, 0, 0): [41.55, 42.15]: [8.1, 8.6]: 62
(0, 0, 1): [41.55, 42.15]: [8.6, 8.9]: 61
(0, 0, 2): [41.55, 42.15]: [8.9, 9.7]: 60
(0, 1, 0): [42.15, 42.33]: [8.4, 8.8]: 62
(0, 1, 1): [42.15, 42.33]: [8.8, 9.0]: 61
(0, 1, 2): [42.15, 42.33]: [9.0, 9.8]: 60
(0, 2, 0): [42.33, 43.20]: [8.6, 8.9]: 61
(0, 2, 1): [42.33, 43.20]: [8.9, 9.2]: 61
(0, 2, 2): [42.33, 43.20]: [9.2, 9.8]: 60
(1, 0, 0): [42.01, 42.42]: [8.0, 8.9]: 62
(1, 0, 1): [42.01, 42.42]: [8.9, 9.1]: 61
(1, 0, 2): [42.01, 42.42]: [9.1, 10.0]: 60
(1, 1, 0): [42.42, 42.56]: [8.2, 9.0]: 61
(1, 1, 1): [42.42, 42.56]: [9.0, 9.2]: 61
(1, 1, 2): [42.42, 42.56]: [9.2, 9.9]: 60
(1, 2, 0): [42.56, 43.45]: [8.7, 9.1]: 61
(1, 2, 1): [42.56, 43.45]: [9.1, 9.3]: 61
(1, 2, 2): [42.56, 43.45]: [9.3, 10.0]: 60
(2, 0, 0): [42.09, 42.59]: [8.1, 9.1]: 62
(2, 0, 1): [42.09, 42.59]: [9.1, 9.2]: 61
(2, 0, 2): [42.09, 42.59]: [9.2, 10.0]: 61
(2, 1, 0): [42.59, 42.74]: [8.7, 9.1]: 62
(2, 1, 1): [42.59, 42.74]: [9.1, 9.3]: 61
(2, 1, 2): [42.59, 42.74]: [9.3, 10.1]: 60
(2, 2, 0): [42.74, 43.53]: [8.6, 9.3]: 62
(2, 2, 1): [42.74, 43.53]: [9.3, 9.5]: 61
(2, 2, 2): [42.74, 43.53]: [9.5, 10.1]: 60
(3, 0, 0): [42.42, 42.78]: [8.7, 9.2]: 61
(3, 0, 1): [42.42, 42.78]: [9.2, 9.4]: 61
(3, 0, 2): [42.42, 42.78]: [9.4, 10.0]: 60
(3, 1, 0): [42.78, 42.90]: [8.6, 9.3]: 61
(3, 1, 1): [42.78, 42.90]: [9.3, 9.4]: 61
(3, 1, 2): [42.78, 42.90]: [9.4, 10.0]: 60
(3, 2, 0): [42.90, 43.62]: [8.8, 9.4]: 61
(3, 2, 1): [42.90, 43.62]: [9.4, 9.6]: 60
(3, 2, 2): [42.90, 43.62]: [9.6, 10.1]: 60
(4, 0, 0): [42.02, 42.97]: [8.5, 9.2]: 62
(4, 0, 1): [42.02, 42.97]: [9.2, 9.5]: 61
(4, 0, 2): [42.02, 42.97]: [9.5, 10.1]: 60
(4, 1, 0): [42.97, 43.13]: [8.7, 9.3]: 61
(4, 1, 1): [42.97, 43.13]: [9.3, 9.5]: 61
(4, 1, 2): [42.97, 43.13]: [9.5, 10.1]: 60
(4, 2, 0): [43.13, 43.92]: [8.8, 9.4]: 61
(4, 2, 1): [43.13, 43.92]: [9.4, 9.7]: 61
(4, 2, 2): [43.13, 43.92]: [9.7, 10.4]: 60
\end{filecontents*}
\pgfplotstabletypeset[
        header=false,
        multicolumn names, % allows to have multicolumn names
        col sep=colon, % the seperator in our .csv file
        display columns/0/.style={
            column name={$\left( z_{\rm gp}, L_{\rm gp}, M_{\rm gp} \right )$}, % name of first column
            string type},
        display columns/1/.style={
            column name={$\log{\text{\lwv}}$},
            string type},
        display columns/2/.style={
            column name={$\log{\text{\mbh}}$},
            string type},
        display columns/3/.style={
            column name={$N_{\rm quasar}$},
            string type},      
        every head row/.style={
            before row={\toprule}, % have a rule at top
            after row={
             & \text{erg/s/\AA} & $M_\odot$ & \\  % the units seperated by &             
            \midrule} % rule under units
            },
        every row no 0/.append style={before row={\multicolumn{4}{c}{Group $z=[0.698 , 0.961]$ with} \\
            \multicolumn{4}{c}{$\log{(\lambda_{\rm rf,o}/\text{\AA})}=[3.54 , 3.60]$, $\log{(\lambda_{\rm rf,c}/\text{\AA})}=[3.43 , 3.50]$} \\
            \midrule}}, 
        every row no 8/.style={after row=\midrule},
        every row no 9/.append style={before row={\multicolumn{4}{c}{Group $z=[0.961 , 1.226]$ with} \\
            \multicolumn{4}{c}{$\log{(\lambda_{\rm rf,o}/\text{\AA})}=[3.48 , 3.54]$, $\log{(\lambda_{\rm rf,c}/\text{\AA})}=[3.38 , 3.43]$} \\
            \midrule}},
        every row no 17/.style={after row=\midrule},            
        every row no 18/.append style={before row={\multicolumn{4}{c}{Group $z=[1.226 , 1.490]$ with} \\
            \multicolumn{4}{c}{$\log{(\lambda_{\rm rf,o}/\text{\AA})}=[3.44 , 3.48]$, $\log{(\lambda_{\rm rf,c}/\text{\AA})}=[3.33 , 3.38]$} \\
            \midrule}},
        every row no 26/.style={after row=\midrule},
        every row no 27/.append style={before row={\multicolumn{4}{c}{Group $z=[1.490 , 1.769]$ with} \\
            \multicolumn{4}{c}{$\log{(\lambda_{\rm rf,o}/\text{\AA})}=[3.39 , 3.44]$, $\log{(\lambda_{\rm rf,c}/\text{\AA})}=[3.28 , 3.33]$} \\
            \midrule}},
        every row no 35/.style={after row=\midrule},
        every row no 36/.append style={before row={\multicolumn{4}{c}{Group $z=[1.769 , 2.400]$ with} \\
            \multicolumn{4}{c}{$\log{(\lambda_{\rm rf,o}/\text{\AA})}=[3.30 , 3.39]$, $\log{(\lambda_{\rm rf,c}/\text{\AA})}=[3.20 , 3.28]$} \\
            \midrule}},                             
        every last row/.style={after row=\bottomrule} % rule at bottom
    ]{Tab1.csv} % filename/path to file
  \end{center}
\end{table}

\subsection{Binning}
\label{sec:bin}

We split the 2\,737 quasars within the redshift range of $0.698 < z < 2.4$ into five redshift groups with roughly the same size in number of quasars. Within each $z$ bin, we further divide the quasars into three \lwv{} groups. Moreover, within each \lwv{} bin, we divide the quasars into three \mbh{} bins. Each ($z$, \lwv, \mbh) group contains a similar number of quasars. The boundaries of $z$, \lwv, and \mbh{} of each bin is shown in Table~\ref{tab:bin}. The SF is evaluated as a function of the intrinsic (rest-frame) time intervals, $\Delta t$, and represented as $\log{A(\Delta t)}$. Within each ($z$, \lwv, \mbh) group as well as each $z$ group of quasars, we divide the observational pairs into 50 $\Delta t$ bins. The shortest bin with $\Delta t<1$ is used for noise analysis and thus forced to $A=0$; the remaining bins are balanced in terms of the number of pairs. This operation is carried out separately for the orange and cyan data. Overall, there are a total of $2\,250$ bins in ($z$, \lwv, \mbh, $\Delta t$) and $300$ bins in ($z$, $\Delta t$) for each passbands. We represent the centre of each $\Delta t$ bin by the average $\log{\Delta t}$ among the pairs in each bin. In total, there are 1.0 and 0.08 billion observational pairs for the orange and cyan passband, respectively, used in the following analysis.

\subsection{Simple bin average structure function}
\label{sec:nobo}

For each $\Delta t$ bin of each quasar, we calculate an SF amplitude, $A$, with Equation~(\ref{eq:sf}) using all available pairs.
We calculate the magnitude differences, \dm, using the 3$\sigma$-clipped mean among the pairs, and the noise, \sigsq, using the second definition in Sec.~\ref{sec:VSF}. Then, for each ($z$, \lwv, \mbh, $\Delta t$) bin and each ($z$, $\Delta t$) bin we calculate the median squared amplitude, $A^2$, and the standard error of the median among quasars, which ensures that quasars with negative noise-corrected amplitudes $A^2$, due to over-subtraction by the \sigsq, are still included in the statistics. Finally, both the amplitude and its error are converted into log space for the following fitting.

\subsection{Fitting}
\label{sec:fit}

We conduct Levenberg-Marquardt least-squares fits \citep{Ma09} to the $A(\Delta t)$ of each ($z$, \lwv, \mbh) and each $z$ group of quasars for two models to examine whether there is a break or not.

The first model is a linear equation:
\begin{equation}
    \log{A}=A_{0,\rm lin}+\text{\glin}\log{\Delta t}.
\end{equation}
The $A_{0,\rm lin}$ is the constant and \glin{} is the slope. We perform two fits under this model, one with a free slope \glin{}, and one with a fixed random-walk slope of \glin$=\gamma_{\rm RW}=0.5$.

The second model is a smoothed piecewise function:
\begin{equation}
    \begin{split}
        \log{A} &=A_{\rm 0,pw}-\frac{\log{\text{\tbrk}}}{2}(\gamma_{\rm 2,pw}-\gamma_{\rm 1,pw})+\gamma_{\rm 2,pw}\log{\Delta t} \\
         &+\frac{(\gamma_{\rm 2,pw}-\gamma_{\rm 1,pw})}{\alpha_{\rm pw}}\ln{(1+e^{-\alpha_{\rm pw}(\log{\Delta t}-\log{\text{\tbrk}})})}.
    \end{split}
	\label{eq:2pie}
\end{equation}
The $A_{\rm 0,pw}$ is the constant, the \tbrk{} is where the $\Delta t$ having the break, the \gsh{} and \glo{} are the slopes at shorter and longer timescales, respectively, and the $\alpha_{\rm pw}$ is the smoothing parameter.
To simplify the comparison with the first model, we fix \tbrk$=30$~day, \gsh$=0.75$, \glo$=\gamma_{\rm RW}=0.5$, and $\alpha_{\rm pw}=5$ in the fitting of the second model. 

The shorter-timescale end of the fitting range is decided based on the purpose while the longer-timescale end is chosen to exclude possible window effects.

\section{Results and discussion}
\label{sec:resul}

\subsection{Break}
\label{sec:brk}

\begin{figure*}
\begin{center}
\includegraphics[width=0.99\textwidth]{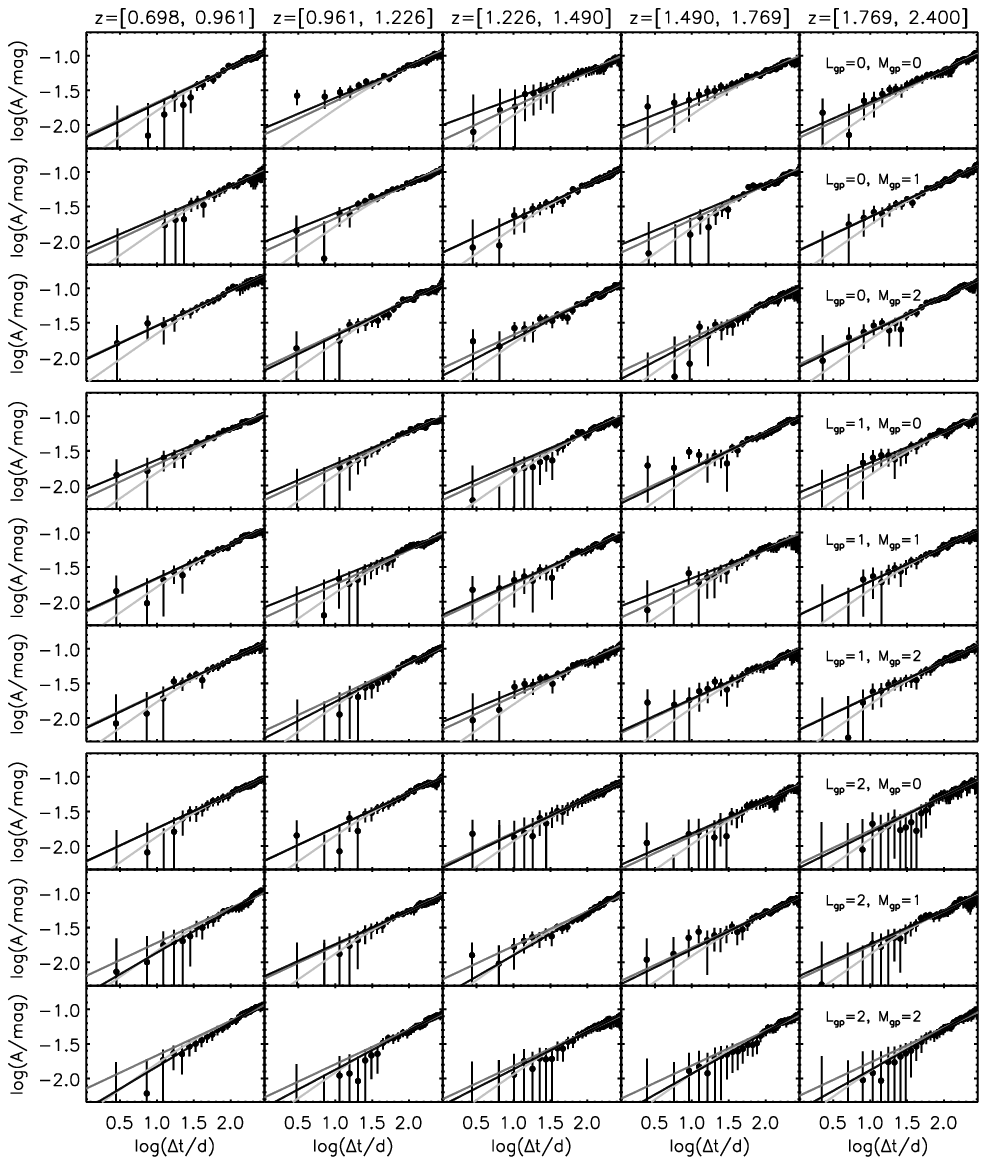}
\caption[$A(\Delta t)$ under ($z$, \lwv, \mbh) grouping method for orange data.]{The $A(\Delta t)$ under ($z$, \lwv, \mbh) grouping method for orange data. Three least-squares fits are shown as lines. Two of them are using the linear model, one with a free slope $\gamma$ (black) and one with $\gamma_{\rm RW}=0.5$ (dark grey). The other one is using the smoothed piecewise model (grey). The fitting range is $0.2<\log{(\Delta t)}<2.4$. The panels are labelled with group indices, and their boundaries in \lwv\ and \mbh{} are shown in Table~\ref{tab:bin}.
}
\label{fig:sf_o}
\end{center}
\end{figure*}

\begin{figure*}
\begin{center}
\includegraphics[width=\textwidth]{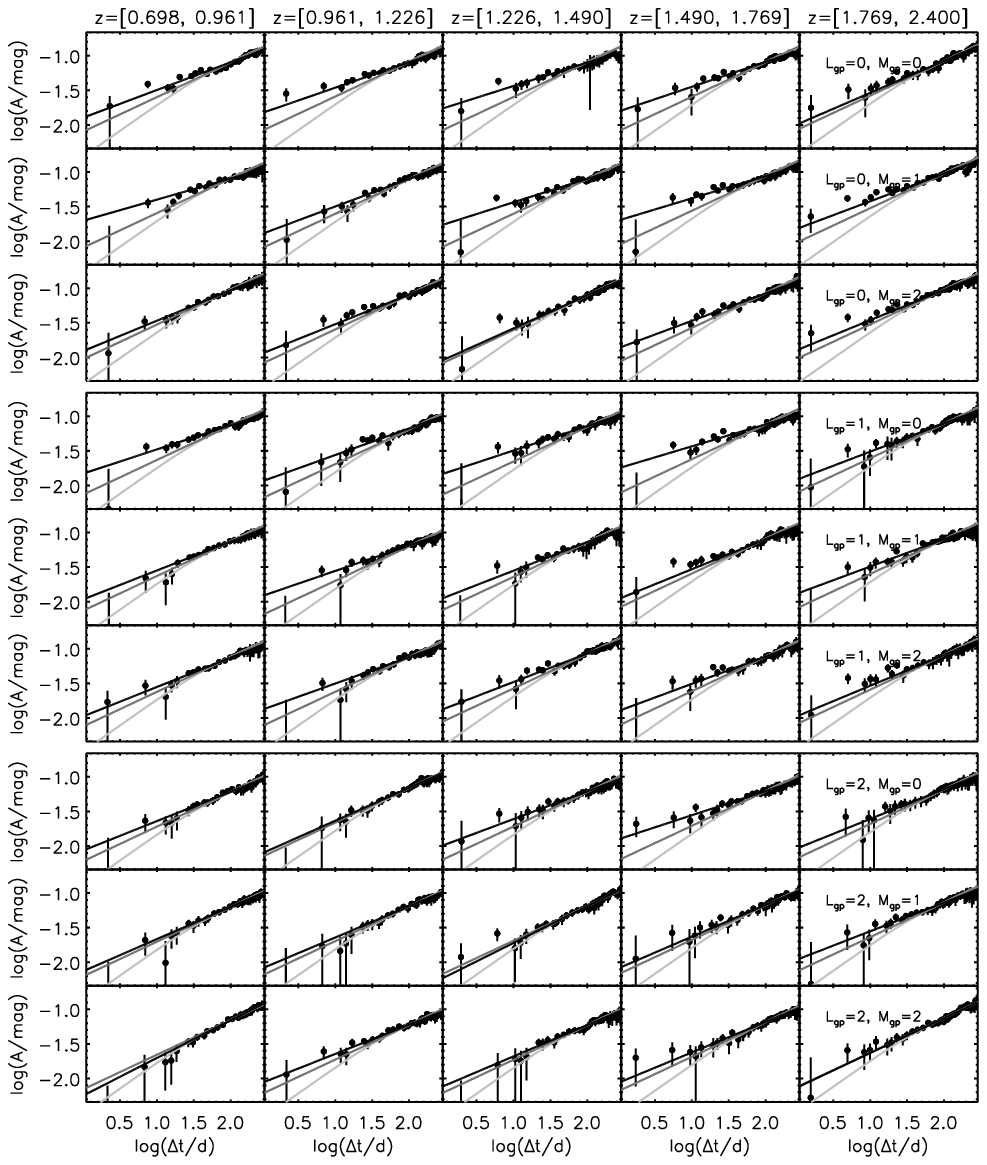}
\caption[The $A(\Delta t)$ under ($z$, \lwv, \mbh) grouping method for cyan data.]{The $A(\Delta t)$ under ($z$, \lwv, \mbh) grouping method for cyan data. The format is the same as in Figure~\ref{fig:sf_o}.
}
\label{fig:sf_c}
\end{center}
\end{figure*}

\begin{figure*}
\begin{center}
\includegraphics[width=\textwidth]{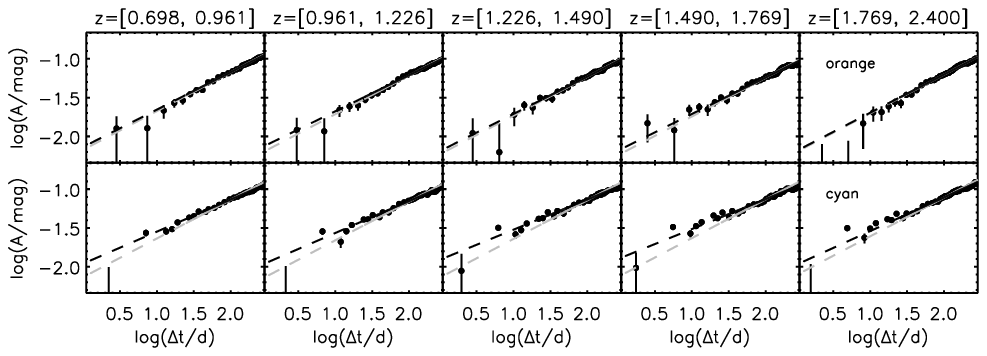}
\caption[The $A(\Delta t)$ under $z$ group only.]{The $A(\Delta t)$ with $z$ grouping only. Two least-squares fits are shown as lines, one with a free slope $\gamma$ (black) and one with $\gamma_{\rm RW}=0.5$ (grey). Lines are solid in the fitting range and dashed in the extrapolation.
}
\label{fig:sf_z}
\end{center}
\end{figure*}

We show the SF result under ($z$, \lwv, \mbh) grouping with their fits of orange and cyan data in Figs.~\ref{fig:sf_o} and~\ref{fig:sf_c}, respectively. We first look at the structure functions across this variety of group to make sure that we are not missing any breaks that may depend on mass or luminosity and get washed out in an overall SF for the whole sample. To determine whether the break exists, we compare the fixed slope linear fit and the smoothed piecewise fit under each ($z$, \lwv, \mbh) group. The fitting range is $0.2<\log{(\Delta t)}<2.4$ so that their performance on the shorter timescale can be considered. Since the two models are nested, i.e., one model can be obtained simply by fixing parameters in the other model (if \gsh$=0.5$ in the smoothed piecewise model, it becomes identical with the fixed slope linear model), we can judge which model is a better one by comparing their $\chi_\nu^2$ values. The model with a significantly smaller $\chi_\nu^2$ is certainly a better fit. If they are comparable, then they are probably equally good. The result shows that most of the groups have comparable $\chi_\nu^2$ values between those two fits in the orange SF, while more than half of the groups have significant better $\chi_\nu^2$ value for the fixed linear model fit in the cyan SF. None of the groups show a significantly better $\chi_\nu^2$ value for the smoothed piecewise model fit. Although this result can not rule out the smoothed piecewise model completely, a simpler model is usually favoured under this circumstance. Therefore, we think the linear function should be sufficient to describe the SF behaviour in our data and no break is needed.

We then try to reduce the noise by constructing fewer groups with more objects by marginalizing over luminosity and mass, and examine the SF with grouping only in redshift, i.e. restframe wavelength range, in Fig.~\ref{fig:sf_z}. The fitting range here is $1.4<\log{(\Delta t)}<2.4$ to capture the main random-walk portion of the SF and exclude any possible short-term suppression that has been reported previously. The dashed line at shorter timescales extrapolates the fit and appears very consistent with the data, suggesting a random-walk relation can describe the data without hints of a break over the full range of $\Delta t=[10;250]$~days. Below that time scale, our results are consistent with a continued random walk of the same slope, but the noise prevents us from claiming more detail. Figure 3 of \citetalias{TWT} had shown breaks on the short timescale, which we now argue are most likely due to over-subtraction of the noise. However, the noise subtraction and short-$\Delta t$ behaviour do not affect their main scientific result on longer timescales ($>1$ month).

Early works suggesting the $\gamma$ slope change from 0.5 to $>1$ with breaks between $5$ to $50$~days were carried out on high-cadence light curves from the Kepler mission for a handful of individual quasars \citep[e.g.][]{Edel14, Sm18}. Further works using ground-based telescopes to obtain LCs for $\sim100$ to $\sim200$ quasars found breaks between $30$ to $300$~days \citep[e.g.][]{Simm16, St22}. We note, that the de-correlation timescale in the DRW model should produce a break that is distinct from a short-timescale break we study here. In theory, it is expected at a much longer timescale and be a transition from a random-walk slope to a potentially flat slope with a saturated variability amplitude instead of a transition from a power-law slope to a steep slope going to shorter time intervals. However, if the structure function does not feature sharp breaks but appears gradually curved, the interpretation of breaks becomes ambiguous \citep[e.g.][]{Ka15a, Arev24}. Our result shows neither breaks nor much curvature, but is consistent with other studies using samples of over $10\,000$ quasars \citep[e.g.][]{Mo14, Ca17}, which show no strong signs of short-term suppression in a break on timescales above a few days. Other than a possible over-subtraction of noise in the SF or PSD leading to short-term breaks, \citet{More21} discovered an unidentified instrumental issue in Kepler data that may affect measurements of stochastic variability beyond repair. Hence, we do not know for sure at this stage where to trust or re-analyse Kepler data on quasar variability. The jury thus seems to be out on where in the AGN parameter space and why breaks in the SF due to a short-term amplitude suppression occur.

Of course, past observations of breaks on short timescale prompted theoretical speculations of their origin. 
E.g., \citet{Ta20} proposed that an amplitude suppression of short-term variability could be caused by a kernel filtering effect. They convolved the PSD with different kinds of kernel functions and successfully reproduced breaks on short timescales. If variability in a quasar disc is synchronized within any given annulus of the disc, and the disc is inclined relative to the line-of-sight, the broad range of light travel time around a given annulus could act as the physical source of the filtering effect; this would then result in a correlation between the breaks and accretion disc size (projected onto the line-of-sight) for different quasars. However, an absence of breaks as in our observations does not imply limits on the sizes of discs, because it is not clear how synchronized any variability is and how strong the expected effect would be.

\begin{figure*}
\begin{center}
\includegraphics[width=\textwidth]{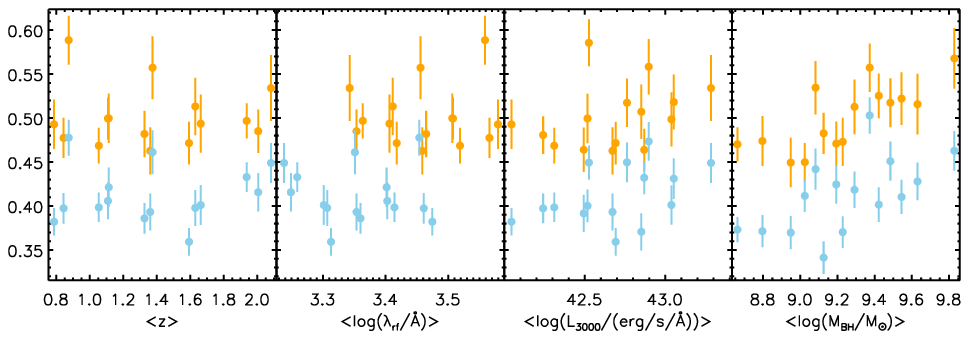}
\caption[]{The slope $\gamma$ of the structure function is shown as a function of quasar properties from Table~\ref{tab:bin}. The points shown here are re-grouped according to their x-loci to reduce the noise in the results of the ($z$, \lwv, \mbh) grouping.  
The mean slopes and their errors are $\gamma_{\rm o}=0.502\pm0.007$ (orange) and $\gamma_{\rm c}=0.412\pm0.005$ (cyan), while their root-mean-square values are $0.056$ (orange) and $0.053$ (cyan).
}
\label{fig:slp}
\end{center}
\end{figure*}

There is strong evidence for breaks on a time scale of a few days in X-ray variability \citep[e.g.][]{Pao23}. While the X-ray emitting region is small enough so that light travel time filtering is not to be blamed, the lamppost model advocates that UV/optical variability is caused by X-ray variability, suggesting that an X-ray break would propagate into a UV-break as well. However, recent magneto-hydrodynamic simulations by \citet{Sec24} demonstrated that X-ray variability is not the principal driver of the UV/optical variability, and the PSD shape of the UV/optical variability is hardly affected by the X-ray behaviour. 

\subsection{Slope}
\label{sec:slp}

Finally, in Fig.~\ref{fig:slp} we look for trends in the best-fitting slope $\gamma$ with quasar properties (Table~\ref{tab:bin}). The fitting range is chosen to be $1.4<\log{(\Delta t)}<2.4$ to avoid any possible short-term suppression. Although most of them are consistent with each other in Figs.~\ref{fig:sf_o},~\ref{fig:sf_c}, and~\ref{fig:sf_z}, we notice that the orange SF generally has slightly steeper slopes than the cyan data. The mean slopes and their errors for orange and cyan data are $\gamma_{\rm o}=0.502\pm0.007$ and $\gamma_{\rm c}=0.412\pm0.005$. Given that we took care to subtract the noise level such that we forced the result to be zero variability on a timescale of rest-frame hours, it is hard to see how this could be due to wrong noise treatment. However, it is the case that groups with more massive \mbh{} seem to have steeper slopes in Fig.~\ref{fig:slp} and also have lower noise levels than groups with less massive black holes. In line with that, no significant differences exist among the noise levels of different $z$ or \lrf \ groups, and no strong dependence of the slope is seen with these parameters either. The challenge for a physical interpretation of different power law slopes in the two passbands is that a fixed restframe wavelength appears in one band for one redshift and the other band for another redshift. This suggests that the slope is affected by an instrumental aspect in the SF analysis even though the noise level should have only a minor effect on a linear scale. This prevents us from further interpreting the slope with any physical properties of quasars. Note, that the slope differences observed here are modest and still statistically consistent with a random walk slope of $0.5$, which is supported by that fact that fixed $\gamma_{\rm RW}=0.5$ fits in Figs.~\ref{fig:sf_o},~\ref{fig:sf_c}, and~\ref{fig:sf_z} are consistent with the data points themselves. 

\section{Conclusions}
\label{sec:conclu}

We use five years of NASA/ATLAS data to study the short timescales quasar variability in two optical passbands, cyan and orange. In our sample, we have more than 2\,700 quasars with the physical properties of $z$, \lwv, and \mbh. According to these properties, we group the quasars into 45 ($z$, \lwv, \mbh) groups and five $z$ groups to perform the structure function analysis. Using an improved definition of the noise level, we show that the model without breaks is sufficient all the way to a short timescale of $\sim$ 10 days, contradicting some of the previous observations and theoretical predictions. Although we can not rule out a piecewise model completely, a simpler model is usually favoured under this circumstance. Our result is consistent with the random walk prediction, suggested by the disc instability model.

\section*{Acknowledgements}

We thank an anonymous referee for suggestions improving the manuscript. JJT was supported by the Taiwan Australian National University PhD scholarship, the Australian Research Council (ARC) through Discovery Project DP190100252, the National Science and Technology Council (MOST 111-2112-M-002-015-MY3), the Ministry of Education, Taiwan (MOE Yushan Young Scholar grant NTU-110VV007, NTU-110VV007-2, NTU-110VV007-3), the National Taiwan University research grant (NTU-CC-111L894806, NTU-CC-112L894806, NTU-CC-113L894806), and also acknowledges support by the Institute of Astronomy and Astrophysics, Academia Sinica (ASIAA). JT has been funded in part by the Stromlo Distinguished Visitor Program at RSAA. We thank I-Non Chiu and Jennifer I-Hsiu Li for suggestions improving the manuscript. This research has made use of \textsc{idl}.

This work uses data from the University of Hawaii's ATLAS project, funded through NASA grants NN12AR55G, 80NSSC18K0284, and 80NSSC18K1575, with contributions from the Queen's University Belfast, STScI, the South African Astronomical Observatory, and the Millennium Institute of Astrophysics, Chile.

This work has made use of SDSS spectroscopic data. Funding for the Sloan Digital Sky Survey IV has been provided by the Alfred P. Sloan Foundation, the U.S. Department of Energy Office of Science, and the Participating Institutions. SDSS-IV acknowledges support and resources from the Center for High Performance Computing at the University of Utah. The SDSS website is \href{http://www.sdss.org}{http://www.sdss.org}. SDSS-IV is managed by the Astrophysical Research Consortium for the Participating Institutions of the SDSS Collaboration including the Brazilian Participation Group, the Carnegie Institution for Science, Carnegie Mellon University, Center for Astrophysics | Harvard \& Smithsonian, the Chilean Participation Group, the French Participation Group, Instituto de Astrof\'isica de Canarias, The Johns Hopkins University, Kavli Institute for the Physics and Mathematics of the Universe (IPMU) / University of Tokyo, the Korean Participation Group, Lawrence Berkeley National Laboratory, Leibniz Institut f\"ur Astrophysik Potsdam (AIP), Max-Planck-Institut f\"ur Astronomie (MPIA Heidelberg), Max-Planck-Institut f\"ur Astrophysik (MPA Garching), Max-Planck-Institut f\"ur Extraterrestrische Physik (MPE), National Astronomical Observatories of China, New Mexico State University, New York University, University of Notre Dame, Observat\'ario Nacional / MCTI, The Ohio State University, Pennsylvania State University, Shanghai Astronomical Observatory, United Kingdom Participation Group, Universidad Nacional Aut\'onoma de M\'exico, University of Arizona, University of Colorado Boulder, University of Oxford, University of Portsmouth, University of Utah, University of Virginia, University of Washington, University of Wisconsin, Vanderbilt University, and Yale University.

This work has made use of data from the European Space Agency (ESA) mission {\it Gaia} (\href{https://www.cosmos.esa.int/gaia}{https://www.cosmos.esa.int/gaia}), processed by the {\it Gaia} Data Processing and Analysis Consortium (DPAC, \href{ https://www.cosmos.esa.int/web/gaia/dpac/consortium}{ https://www.cosmos.esa.int/web/gaia/dpac/consortium}). Funding for the DPAC has been provided by national institutions, in particular the institutions participating in the {\it Gaia} Multilateral Agreement.

%%%%%%%%%%%%%%%%%%%%%%%%%%%%%%%%%%%%%%%%%%%%%%%%%%
\section*{Data Availability}
The data underlying this article will be shared on reasonable request to the corresponding author. 
%The inclusion of a Data Availability Statement is a requirement for articles published in MNRAS. Data Availability Statements provide a standardised format for readers to understand the availability of data underlying the research results described in the article. The statement may refer to original data generated in the course of the study or to third-party data analysed in the article. The statement should describe and provide means of access, where possible, by linking to the data or providing the required accession numbers for the relevant databases or DOIs.



%\begin{thebibliography}{99}
%\end{thebibliography}

%%%%%%%%%%%%%%%%%%%%%%%%%%%%%%%%%%%%%%%%%%%%%%%%%%

%%%%%%%%%%%%%%%%% APPENDICES %%%%%%%%%%%%%%%%%%%%%

%\onecolumn

\begin{appendix}

\end{appendix}

%%%%%%%%%%%%%%%%%%%%%%%%%%%%%%%%%%%%%%%%%%%%%%%%%%

% Don't change these lines
\bsp	% typesetting comment
\label{lastpage}
\end{document}